\documentclass{article}
\usepackage[a4paper]{geometry}
\usepackage{amssymb,amsmath,amsthm,tikz,mathtools,hyperref}
\usetikzlibrary{shapes,arrows}
\tikzstyle{every loop}=[]

\newcommand{\openone}{\leavevmode\hbox{\small1\kern-3.8pt\normalsize1}}

\newcommand{\be}{\begin{equation}}
\newcommand{\ee}{\end{equation}}

\theoremstyle{plain}

\newtheorem{definition}{Definition}

\newcommand{\bigp}{\mathbf{P}}
\newcommand{\p}{\mathbf{p}}
\newcommand{\RT}{\mathcal{R_T}}
\newcommand{\IRT}{\mathcal{\widetilde{R}_T}}
\newcommand{\theory}{\mathcal{T}}
\newcommand{\see}{C}

\begin{document}
\title{Abstraction/Representation Theory for Heterotic Physical Computing}
\author{Dominic C. Horsman\footnote{DCH published previously as Clare Horsman.} \\
\small{Department of Computer Science, University of Oxford, Parks Road, Oxford OX1 3QD}}
\maketitle

\begin{abstract}
We give a rigorous framework for the interaction of physical computing devices with abstract computation. Device and program are mediated by the non-logical \textit{representation relation}; we give the conditions under which representation and device theory give rise to commuting diagrams between logical and physical domains, and the conditions for computation to occur. We give the interface of this new framework with currently existing formal methods, showing in particular its close relationship to refinement theory, and the implications for questions of meaning and reference in theoretical computer science. The case of hybrid computing is considered in detail, addressing in particular the example of an internet-mediated \textit{social machine}, and the abstraction/representation framework used to provide a formal distinction between heterotic and hybrid computing. This forms the basis for future use of the framework in formal treatments of nonstandard physical computers.
%In this paper we introduce computer scientists to AR theory. Abstract Abstract Abstract Abstract Abstract Abstract Abstract Abstract Abstract Abstract Abstract Abstract Abstract Abstract Abstract Abstract Abstract Abstract Abstract Abstract Abstract Abstract Abstract Abstract Abstract Abstract Abstract Abstract Abstract Abstract Abstract Abstract Abstract Abstract Abstract Abstract Abstract Abstract 
\end{abstract}

\section{Introduction}

\begin{quote} \textit{Computer Science is no more about computers than astronomy is about telescopes. } \\
-- E. Dijkstra (attrib.) \end{quote}

\begin{quote} \textit{That is incorrect. } \\
-- Lt. Cmdr. Data (Star Trek)\end{quote}

\noindent Heterotic hybrid computing involves the combination of two or more distinct computational substrates into a computing system whose power is greater than that of its individual components \cite{SS-UC11,susanvivspecis}. Such devices have come into prominence with the slowdown of Moore's Law, and the subsequent search for different ways to increase computing power beyond standard silicon-based digital computers. The most well-developed such system is quantum computing \cite{Ladd2010}; while not usually presented as a hybrid device, a realistic quantum computer requires a large and integrated classical control computer, in particular for performing error correction \cite{blueprint}. Other examples include proposed chemical \cite{chem1,chem2} and biological \cite{dna1,dna2} nonstandard computers, again requiring classical control. Also included are biological systems interfacing with digital software (for instance \cite{ratonaplane}). A different direction for heterotic systems is opened up by the enormous data gathering and communication possibilities available online. A key recent insight there has been the recognition of `social machines': computational ecosystems comprising both digital computers and multiple human users, all acting towards a computational goal \cite{sociam1,sociam2}. Important social machines include Wikipedia \cite{wiki} and crowdsourced science such as Galaxy Zoo \cite{galaxy}. As the space of these devices and interactions expands, heterotic forms of computing will continue to be a growth area in the very near future.

With the expansion of importance of such systems, questions of their computational foundations become newly prominent. Discussion within the unconventional computing community has often become stuck in discussion over (super-)Turing computable behaviour. For practical hybrid devices there are much more pressing concerns. The large toolkit of mathematical and abstract methods for computer science (semantics, logical calculi, programming, verification, etc) has historically been developed in parallel with standard digital computers. Unfortunately, non-standard computing substrates and devices almost never admit a native description in terms of such standard abstract methods. This means that their computing power is both largely unknown and underutilized. They also lack the formal underpinnings that have proved so powerful in standard computer science. For hybrid devices in particular, a lack of formalisation of the individual components means a lack of a combined formalisation when they are put together.

The lack of formalism for novel devices stems not (just) from a lack of understanding about their abilities, but from the way in which standard formal methods have been developed and used. The groundbreaking understanding of Lovelace, Turing, and Shannon, which gave rise to modern computer science, was that the computers they were considering had a theoretical representation separate from their specific physical implementation. While inspired by certain physical devices (the Analytical Engine, the Bombe, the differential analyzer, etc), the associated formal systems could be manipulated without reference to any physically realised system. The interface between the abstract methods of computer science and the physical engineering of computers was, however, not itself formalised. Computers appear within the formalism, if at all, as mathematical or logical abstractions, not as the physical devices themselves.

 This separation proved to be extremely powerful, and was largely responsible for the rapid expansion of the modern science of computers. However as nonstandard, and in particular hybrid, devices now come to prominence, having only an informal connection between the physical device and the abstract formal theory becomes significantly problematic. With a purely informal interface between the two, there is no straightforward way to reach standard abstract computer science from a novel computing substrate. Even if an ad-hoc mapping can be considered for some particular non-standard devices (e.g.. a certain neural net `implements an {\sc xor}'), this cannot be straightforwardly combined with another system with a different mapping to create a hybrid computational device. There is no way formally to combine such informal mappings, meaning that until now there has been no integrated framework that allows for the abstract treatment of nonstandard and hybrid computers.

In this paper we give formal foundations for the interaction of hybrid computing systems through the use of \textit{Abstraction/Representation Theory\footnote{I am indebted to Aleks Kissinger for the name.}} (AR theory). Recently introduced, AR theory allows us make rigorous the interface between physical computing systems and their abstract, mathematical, representation. AR theory was developed in \cite{compute} to address the question of which physical dynamical systems are properly represented as computing systems. In the present paper we give a formal treatment of the theory in the context of the foundations of computer science, centering on the \textit{representation relation} between data processing physical systems and the abstract objects of standard computing theory. AR theory comes equipped with both an algebraic and a graphical language of commuting diagrams, and we show how this interfaces with similar, known, formal methods. The discussion is conducted with particular reference to three elements in the foundations of computer science -- not as an exhaustive list, but to give the reader a broader understanding of the place AR theory occupies in computing theory. We show how AR theory supplements \textit{refinement and reification}, allowing upwards/downwards simulation diagrams to be embedded in physical computing devices. We touch on the category-theoretic underpinnings of such diagrams, and consider also the relationship to the interactions between concrete and abstract semantics given by \textit{Abstract Interpretation}. By grounding computer science in the physical world, questions of ontology and semantics for computational formal systems can also be addressed, and we will focus in particular on \textit{lambda calculi} in this regard. %Finally, AR theory is used to give a formal framework for hybrid computing devices, allowing reasoning across different physical computational substrates. 

In the presence of AR theory, computer science becomes the natural science of the computing abilities of physical systems. While seemingly a very different foundational view of computing, it is in fact a supplement to, rather than a repudiation of, standard computability theory. It is a theory of comput\textit{ing} to add to our current theories of comput\textit{ation}. AR theory allows more physical devices, both individually and as heterotic systems, to be brought under the umbrella of abstract computation theory, while remaining agnostic about questions around their universality and/or Turing completeness. The addition of AR theory to the standard toolkit of computer science promises to greatly increase our power to reason about complex computational physical devices.

\section{Basic elements of AR theory}

AR theory was introduced in \cite{compute}, with that paper giving the full physical and philosophical background for the framework. The present paper is concerned with its formalisation for use specifically in computer science. That previous work should be referenced for questions concerning the broad ontology and context of AR theory, which will not in general be reiterated here.

AR theory concerns objects in the domain of physical systems, abstract objects (including mathematical and logical entities), and the representation relation which mediates between the two. The distinction between the two spaces, abstract and physical, is fundamental in the theory, as is their connection \textit{only} by the (directed) representation relation. An intuitive example is given in figure \ref{reprel}: a physical switch is represented by an abstract bit by some representation.

\begin{figure}[t]
    \scalebox{0.75}{\hspace{1cm}%%%%%%%%%%%%%
%  FIG 1
%%%%%%%%%%%%%

 \begin{minipage}[c]{1.0\linewidth}
  %\centering
  \[
   \begin{array}{ccc}
   \begin{tikzpicture}[font=\large]

\draw[style=dashed] (0,0) -- (6,0);
\draw (0,0.5) node {$Abstract$};
\draw (0,-0.5) node {$Physical$};

%\node[circle,draw] at (3,-1.5) {$e^-$};
\draw[rounded corners,rotate around={45:(3,-1.75)}] (3,-1.75) rectangle (4,-1.55);
\draw[rounded corners,dashed] (3,-1.75) rectangle (4,-1.55);

\draw (3.5,1.25) node {$(0,1)$};

\end{tikzpicture}
& \qquad \qquad \qquad
\quad &
\begin{tikzpicture}[font=\large]

\draw[style=dashed] (0,0) -- (6,0);
\draw (0,0.5) node {$Abstract$};
\draw (0,-0.5) node {$Physical$};

%\node[circle,draw] at (3,-1.5) {$e^-$};
\draw[rounded corners,rotate around={45:(3,-1.75)}] (3,-1.75) rectangle (4,-1.55);
\draw[rounded corners,dashed] (3,-1.75) rectangle (4,-1.55);

\draw (3.5,1.25) node {$(0,1)$};

\draw[->,double] (3.5,-0.65) -- (3.5,.8);
\draw (4,0.25) node {$\mathcal{R}$};

\end{tikzpicture}
\\{}\\
 \mathrm{(a)} & \qquad \quad  &  \mathrm{(b)}  \end{array}
\]
\end{minipage}}
\caption{Basic representation. (a) Spaces of abstract and physical objects (here, a switch with two settings and a binary digit). (b) The directed representation relation $\mathcal{R}$ mediating between the spaces.} \setcounter{figure}{1}\label{reprel}
\end{figure}
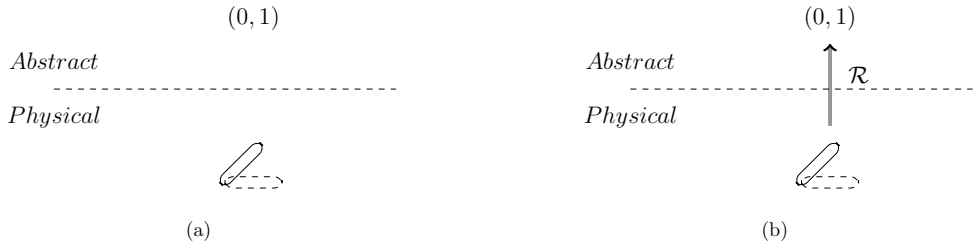

This separation of physical and abstract domains necessitates our first definitions:
\begin{definition} The \textbf{physical domain}, $\bigp$, consists of all physical objects, $\p \in \bigp$.\end{definition}
\begin{definition} The \textbf{abstract domain}, $M$, consists of all abstract objects, $m \in M$. \end{definition}

A \textit{computer} is an object in the domain of physical entities, usually with internal degrees of freedom and physical evolution occurring between initial and final output states. A \textit{computation} is a set of objects and relations within the domain of abstract entities, as described in the logical formalisms of theoretical computer science. Bold font is used to indicated where an object or evolution is physical. Abstract objects are represented using an italic font. 

\begin{figure}[t]
    \scalebox{0.65}{\hspace{-1cm}%%%%%%%%%%%%%
%  FIG 2
%%%%%%%%%%%%%

 \begin{minipage}[c]{1.0\linewidth}
  %\centering
  \[
   \begin{array}{ccc}
   \begin{tikzpicture}[font=\large]

\draw[style=dashed] (0,0) -- (10,0);
\draw (0,0.5) node {$\scriptstyle{Abstract}$};
\draw (0,-0.5) node {$\scriptstyle{Physical}$};

%fig a
\node[draw] (p) at (3,-1) {$\mathbf{p}$};
\node[draw] (mp) at (3,2) {$m_{\mathbf{p}}$};
\draw[->,double] (p) -- (mp);
\draw (3.8,0.5) node {$\RT$};

\end{tikzpicture}

& \qquad  
\quad &   

   \begin{tikzpicture}[font=\large]

\draw[style=dashed] (0,0) -- (10,0);
\draw (0,0.5) node {$\scriptstyle{Abstract}$};
\draw (0,-0.5) node {$\scriptstyle{Physical}$};

%fig a
\node[draw] (p) at (3,-1) {$\mathbf{p}$};
\node[draw] (mp) at (3,2) {$m_{\mathbf{p}}$};
\draw[->,double] (p) -- (mp);
\draw (3.8,0.5) node {$\RT$};

%fig b
\node[draw] (mprp) at (6.5,2) {$m^\prime_{\mathbf{p}}$};
\draw[-open triangle 45] (mp) -- (mprp);
\draw (4.8,2.5) node {$C_\theory(m_\mathbf{p})$};

\end{tikzpicture}
   \\{}\\
 (a) & \qquad \quad  &  (b)  
   \\{}\\{}\\
    \begin{tikzpicture}[font=\large]

\draw[style=dashed] (0,0) -- (10,0);
\draw (0,0.5) node {$\scriptstyle{Abstract}$};
\draw (0,-0.5) node {$\scriptstyle{Physical}$};

%fig a
\node[draw] (p) at (3,-1) {$\mathbf{p}$};
\node[draw] (mp) at (3,2) {$m_{\mathbf{p}}$};
\draw[->,double] (p) -- (mp);
\draw (3.8,0.5) node {$\mathcal{R}_\theory$};

%fig b
\node[draw] (mprp) at (6.5,2) {$m^\prime_{\mathbf{p}}$};
\draw[-open triangle 45] (mp) -- (mprp);
\draw (4.8,2.5) node {$C_\theory(m_\mathbf{p})$};

%fig c
\node[draw] (ppr) at (8,-1) {$\mathbf{p}^\prime$};
\draw[-open triangle 45] (p) -- (ppr);
\draw (5.25,-.75) node {$\mathbf{H}(\mathbf{p})$};

\end{tikzpicture}

& \qquad  
\quad &   

   \begin{tikzpicture}[font=\large]

\draw[style=dashed] (0,0) -- (10,0);
\draw (0,0.5) node {$\scriptstyle{Abstract}$};
\draw (0,-0.5) node {$\scriptstyle{Physical}$};

%fig a
\node[draw] (p) at (3,-1) {$\mathbf{p}$};
\node[draw] (mp) at (3,2) {$m_{\mathbf{p}}$};
\draw[->,double] (p) -- (mp);
\draw (3.8,0.5) node {$\RT$};

%fig b
\node[draw] (mprp) at (6.5,2) {$m^\prime_{\mathbf{p}}$};
\draw[-open triangle 45] (mp) -- (mprp);
\draw (4.8,2.5) node {$C_\theory(m_\mathbf{p})$};

%fig c
\node[draw] (ppr) at (8,-1) {$\mathbf{p}^\prime$};
\draw[-open triangle 45] (p) -- (ppr);
\draw (5.25,-.75) node {$\mathbf{H}(\mathbf{p})$};

%fig d
\node[draw] (mppr) at (8,1) {$m_{\mathbf{p}^\prime}$};
\draw[->,double] (ppr) -- (mppr);
\draw (8.8,-0.3) node {$\RT$};

\end{tikzpicture}
   \\{}\\
(c) & \qquad \quad  &  (d)
 \end{array}
\]
\end{minipage}}
\caption{Parallel evolution of abstract evolution (e.g.. an algorithm) and potential physical computing device. (a) The basic representational triple, $\langle \p , \mathcal{R} , m_\p \rangle$: physical system $\p$ is represented abstractly by $m_\p$ using the modelling representation relation $\RT$ of theory $\theory$. (b) Abstract dynamics $\see_\theory (m_\p)$ give the evolved abstract state $m^\prime_\p $. (c) Physical dynamics $\mathbf{H}(\p)$ give the final physical state $\p^\prime$. (d) $\RT$ is used again to represent $\p^\prime$ as the abstract output $m_{\p^\prime}$.}\label{physabst}
\end{figure}
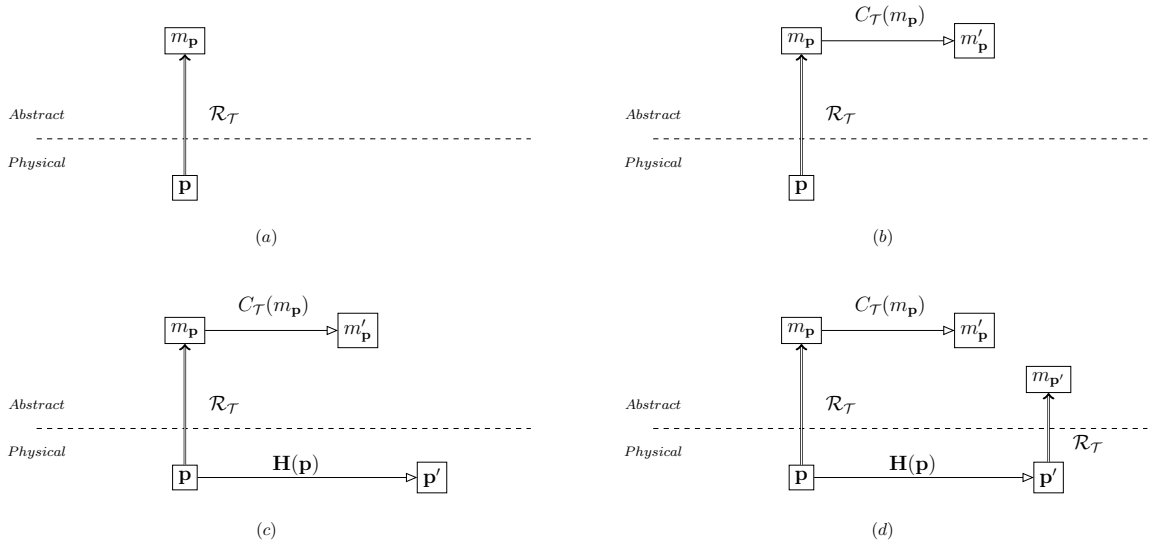

The elementary \textit{representation relation} is the directed map $\mathcal{R} : \bigp \rightarrow M$. When two objects are connected by $\mathcal{R}$ we write them as $\mathcal{R} : \p \rightarrow m_\p$. The abstract object $m_\p$ is then said to be the \textit{abstract representation} of the physical object $\p$, and together they form one of the basic composites of AR theory, the \textit{representational triple} $\langle \p , \mathcal{R} , m_\p \rangle$. The basic representational triple is shown in figure \ref{physabst}(a).

The next map we introduce is that of abstract evolution. This takes abstract objects to abstract objects: $C: M \rightarrow M$. An individual example is shown in figure \ref{physabst}(b), for the mapping $C(m_\p)$ taking $m_\p \rightarrow m'_\p$. The corresponding physical evolution map is given by $\mathbf{H}: \bigp \rightarrow \bigp$. For individual elements in figure \ref{physabst}(c) this is $\mathbf{H(p)}$ which takes $\p \rightarrow \p'$.

In order to reach the next key concept in AR theory, we now apply the representation relation to the outcome state of the physical evolution to give its abstract representation $m_{\p'}$. This forms another representational triple $\langle \p' , \mathcal{R} , m_{\p'} \rangle$, figure \ref{physabst}(d). We now have two abstract object, $m'_\p$ and $m_{\p'}$. For some error quantity $\varepsilon$ and norm $||$, if $|m_{\p'} - m'_{\p}| \le \varepsilon$ then the diagram \ref{physabst}(d) \textit{commutes}. Commuting diagrams are fundamental to the use of AR theory. Definitionally,

\begin{definition} A \textbf{commuting diagram} in AR theory comprises two representational triples $\langle \p , \mathcal{R} , m_\p \rangle$ and $\langle \p' , \mathcal{R} , m_{\p'} \rangle$, and pair of abstract and physical evolutions $C(m_\p): m_\p \rightarrow m'_\p$ and $\mathbf{H(p)}: \p \rightarrow \p'$,  which satisfy the condition $|m_{\p'} - m'_{\p}| \le \varepsilon$.
\end{definition}

If a set of abstract and physical objects form a commuting diagram under representation, then $m_\p$ is a \textit{faithful abstract representation} of physical system $\p$ for the evolutions $C(m_\p)$ and $\mathbf{H(p)}$.

The main implication of having a faithful abstract representation for a physical system is that the final state of a physical object undergoing evolution can be known either by tracking the physical evolution and then representing the output abstractly, or by theoretically evolving the representation of the system. In the first case, the `lower path' of a commuting diagram is followed; in the latter, the `upper path'. Finding out which diagrams commute is the business of basic experimental science, and their initial exploitation that of fundamental engineering and technology. 

% figuring out if diagrams commute is the business of basic science and fundamental technology
In experimental science, a test for commutation of a diagram involves producing a controlled physical setup (the experiment) about which has both an abstract representation $\mathcal{R}$ and an abstract prediction of how it will behave, $C$. The physical system $\mathbf{p}$ is evolved under the physical experimental dynamics $\mathbf{H}$, and the outcome compared to the theoretical prediction. If they coincide within the error tolerance of the experiment and the desired outcome confidence, then the diagram commutes.

This is not, of course, the \textit{purpose} of an experiment. Experiments are designed in order to test not a single scenario but a \textit{theory}. We now bring in the final element of basic AR theory:

\begin{definition}  A \textbf{theory}, $\mathcal{\mathcal{T}}$, is a set of representation relations $\mathcal{R_\mathcal{T}}$ for physical objects, a domain of such objects for which it is purported to be valid, and a set of abstract predictive dynamics for the output of the representations, $m_\p$, $C(m_\p)$.
\end{definition} 

If a theory supports commuting diagrams for all scenarios in which it has been both defined and tested, then it is a \textit{valid} theory\footnote{It may not, however, necessarily be a \textit{good} theory.}.

A physical system or device that is both well tested and well understood will in general have a large number of commuting diagrams supporting it. This is a necessary condition for a theory to be \textit{good}, but not a sufficient one. Furthermore, if a theory is a \textit{good} theory then we can be confident that it will give rise to commuting diagrams outside the domain either or objects or dynamics for which it has been tested.

\section{What representation is not}

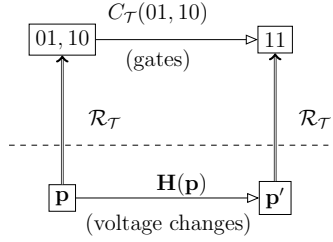
\begin{figure}[t]
    \scalebox{0.7}{\hspace{2cm}%%%%%%%%%%%%%
%  FIG 5
%%%%%%%%%%%%%

 \begin{minipage}[c]{1.0\linewidth}
  %\centering
  \[     
   \begin{tikzpicture}[font=\large]
   
   \draw[style=dashed] (2,0) -- (8,0);
   
   \node[draw] (p) at (3,-1) {$\p$};
\node[draw] (mp) at (3,2) {$01,10$};
\draw[->,double] (p) -- (mp);
\draw (3.8,0.5) node {$\RT$};

\node[draw] (mprp) at (7,2) {$11$};

\draw[-open triangle 45] (mp) -- (mprp);
\draw (4.8,2.5) node {$C_\theory (01,10)$};
\draw (4.8,1.6) node {(gates)};

\node[draw] (ppr) at (7,-1) {$\p^\prime$};
\draw[-open triangle 45] (p) -- (ppr);
\draw (5.25,-.75) node {$\mathbf{H}(\p)$};
\draw (5,-1.5) node {(voltage changes)};

\draw[->,double] (ppr) -- (mprp);
\draw (7.75,0.5) node {$\RT$};

\end{tikzpicture}
 \]
\end{minipage}}
\caption{A commuting diagram for a sequence of gate operations (here, binary addition) for abstract and physical systems, for later use as a computing device.}\label{com}
\end{figure}

Computing relies on having a good and valid well-developed theory of the physical computing device. Commuting diagrams such as figure \ref{com} do not, however, describe the actual process of computing itself, rather are the prerequisites for us to use the physical system as a computer. A computer must, however, satisfy a set of these diagrams. It is therefore useful at this point to consider such commuting diagrams for a computer, and specify exactly what is being described in the AR description of such a device.

A commuting diagram in the context of computation connects the physical computing device, $\p$, and its abstract representation $m_\p$. $m_\p$ can be a number of different abstract representations -- we go into this in more detail below, for now we take $m_\p$ to be drawn from the set of binary strings (see figure  \ref{com}). The abstract evolution is then the (binary) program to be run on the computer, and the physical evolution is how the state of the computer changes during the program (change of voltages etc). The full commuting diagram describes the parallel evolution of physical computer and abstract algorithm. They are connected via the representation given by the theory of the computing device, $\mathcal{R_\mathcal{T}}$. 

What precisely does the relationship given by AR theory between computer and computation consist in? That is, how does AR theory describe and define the relation between physical and abstract computational objects and evolutions, and what attributes does it give to this relationship? It is instructive at this point to compare the AR diagram figure \ref{com} with an existing formal framework that can appear at first sight to be very similar. The representation of computation given by Abstract Interpretation \cite{absint} involves a concrete semantics modelling the operation of the computing system, an abstract semantics representing the algorithm or program to be run, and the functional \textit{abstraction relation} mapping between the two. Is this not, then, precisely what we have described with AR theory?

The reason why the answer to this question is `no' gets to the heart of AR theory and what it is modelling in the representation relation. In Abstract Interpretation, the concrete semantics is designed to represent the operation of the physical device, but is \textit{not the physical device}. It is an abstract, mathematical representation of the physical system, not the physical system itself. As a consequence, the abstraction relation between concrete and abstract semantics is a mathematical function, taking mathematical objects as both its domain and range.

By contrast, reference to the physical system within AR theory is to \textit{the physical system itself}, not a representation of it. $\mathbf{p}$ is a physical object, $m_\p$ its mathematical/logical representation. The concrete semantics of Abstract Interpretation are objects in the space $M$ not the space $\mathbf{P}$. AR theory concerns how physical objects \textit{qua} physical objects relate to their own abstract representation, rather than how different types of abstract representation interact with each other. As a consequence, $\p$ must stand for the physical system and not some pre-defined representation; a model, a program, or a semantics, however low-level, is still a \textit{representation} of the physical system, $m_\p$, not the physical system, $\p$, itself.

This is the core of AR theory, and it cannot be over-emphasised: \textit{$\mathcal{R}$ as a representation relation is \textbf{not} a mathematical relation}. Neither is it a logical relation. It is a relation whose domain is physical objects and whose range is abstract/mathematical objects: it is a \textit{representation} relation. This is an entirely new kind of relation, one that is key to understanding how physical computing devices operate by mediating between the level of physics and the level of data manipulation.

%
%check notation
% abstract and physical evolutions

% Abstract Interpretation
% a model or a semantics is not a physical object
% P = {p} is actual physical stuff - we will come back to this again and again through the paper

%  more about R
% include tau

\section{Computing in AR theory}

The parallel evolution of physical device and abstract program in figure  \ref{com} is not yet the AR representation of using that device to perform a computation. In order to describe computing, we need one final element of AR theory, this time not a basic element but a composed one.

The representation relation defined here is directed, from physical to abstract objects. This is \textit{modelling}: giving an abstract representation of a physical object. The question can now be posed: is it possible to give a reversed representation relation, an \textit{instantiation} relation? This will not be a basic relation in the same way as the ordinary (modelling) representation relation is basic: abstract representation can be given for any physical object (this is language), but there are plenty of abstract objects that do not have a physical instantiation (`unicorn', `infinite-tape Turing machine', $f(x) = \sum_k (\pi k^2)^{-1}\sin (\pi k^2 x)$, etc). Only in very specific circumstances can an instantiation relation $\IRT$ be given for a theory $\mathcal{T}$.

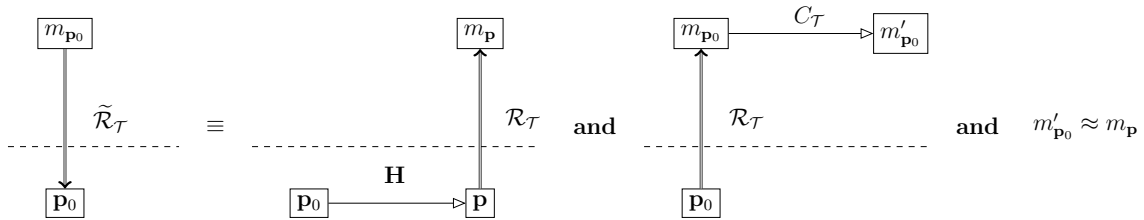
\begin{figure}[t]
    \scalebox{0.75}{\hspace{0cm}%%%%%%%%%%%%%
%  FIG 4.5
%%%%%%%%%%%%%

 \begin{minipage}[c]{1.0\linewidth}
  %\centering
  \[
   \begin{array}{ccccccc}
   \begin{tikzpicture}[font=\large]
   
   \draw[style=dashed] (2,0) -- (5,0);

\node[draw] (p) at (3,-1) {$\p_0$};
\node[draw] (mp) at (3,2) {$m_{\p_0}$};
\draw[->,double] (mp) -- (p);
\draw (3.8,0.5) node {$\IRT$};

\end{tikzpicture}

   & \begin{tikzpicture}[font=\large]
   \draw(0,0) node {};
    \draw(0,1.5) node {$\equiv$};  
    \end{tikzpicture}
     &

      \begin{tikzpicture}[font=\large]
   
   \draw[style=dashed] (2,0) -- (7,0);

\node[draw] (p) at (3,-1) {$\p_0$};
\node[draw] (ppr) at (6,-1) {$\p$};
\node[draw] (mppr) at (6,2) {$ m_{\p}$};

\draw[->,double] (ppr) -- (mppr);
\draw (6.75,0.5) node {$\RT$};

\draw[-open triangle 45] (p) -- (ppr);
\draw (4.5,-0.5) node {$\mathbf{H}$};

\end{tikzpicture}
   
      &  \begin{tikzpicture}[font=\large]
   \draw(0,0) node {};
    \draw(0,1.5) node {\textbf{and}};  
    \end{tikzpicture} &
      
            \begin{tikzpicture}[font=\large]
   
   \draw[style=dashed] (2,0) -- (7,0);

\node[draw] (p) at (3,-1) {$\p_0$};
\node[draw] (mp) at (3,2) {$m_{\p_0}$};
\draw[->,double] (p) -- (mp);
\draw (3.8,0.5) node {$\RT$};

\node[draw] (mprp) at (6.5,2) {$m^\prime_{\p_0}$};
\draw (4.9,2.3) node {$C_\theory$};
\draw[-open triangle 45] (mp) -- (mprp);

\end{tikzpicture}
     
      &  \begin{tikzpicture}[font=\large]
   \draw(0,0) node {};
    \draw(0,1.5) node {\textbf{and}};  
    \end{tikzpicture} &
    
          \begin{tikzpicture}[font=\large]
   \draw(0,0) node {};
    \draw(0,1.5) node {$m^\prime_{\p_0}\approx m_{\p}$};  
    \end{tikzpicture} 

 \end{array}
\]
\end{minipage}}
\caption{The instantiation relation. The combination of the conditions is a commuting diagram.}\label{INST}
\end{figure}

To find these circumstances, consider again the `upper' and `lower' paths of a commuting diagram, $(\p_0 \rightarrow m_{\p_0} \rightarrow m'_{\p_0})$ and $(\p_0 \rightarrow \p \rightarrow [m_{\p} = m'_{\p_0}])$ respectively. Between them, these paths describe the process of finding some $\p_0$ such that when it is subjected to the physical process $\mathbf{H}: \p_0 \rightarrow \p$ it becomes the physical system $\p$ whose abstract representation is $m_\p$. In other words, if both paths are present and form a commuting diagram, the theory $\mathcal{T}$ can be used to \textit{engineer} system $\p$ from system $\p_0$ given a desired abstract specification $m_\p$. Together, then, they stand as a shorthand for the instantiation relation $\IRT$, as in figure \ref{INST}.

The instantiation relation is given only when $\mathcal{T}$ is both good and valid. It then permits the construction of an \textit{instantiational triple}, $\langle m_\p , \IRT , \p \rangle$. The system $\p$ is then the \textit{physical instantiation} of the abstract object $m_\p$ for theory $\mathcal{T}$.

A use of the instantiation relation can be seen as a counterfactual use of the representation relation: which physical system, when represented abstractly, would give the abstract representation that we are trying to instantiate? The method by which it is achieved will vary considerably given different scenarios: trial and error, abstract reasoning, numerical simulation, etc. What connects these methods is that they are not straightforward: it is generally a skillful and creative process to reverse a representation relation.

%\begin{definition} A \textbf{commuting diagram} in AR theory comprises two representational triples $\langle \p , \mathcal{R} , m_\p \rangle$ and $\langle \p' , \mathcal{R} , m_{\p'} \rangle$, and pair of abstract and physical evolutions $C(m_\p): m_\p \rightarrow m'_\p$ and $\mathbf{H(p)}: \p \rightarrow \p'$,  which satisfy the condition $|m_{\p'} - m'_{\p}| \le \varepsilon$.

% reversed RR
% engineering
%AR compute cycle

In computing, the use of an instantiation relation is the act of \textit{encoding and initialisation}. At its simplest, this is the encoding of abstract data in a physical device, from turning a dial to a specified input state, punching a set of holes on a card, to initialising a series of voltages across a semiconductor. In all cases, how data are represented by physical objects is determined both by the available physics of the system and by design choices.  %different examples. choice and physics determines is

Initialisation is the first step in the AR cycle for computing. This has one immediate implication: that a device cannot be used as a computer until its theory, $\mathcal{T}$, is well-understood, as good and valid. If it is not, then $\IRT$ cannot be formed with any confidence, and only experiment or engineering or technology cycles can be constructed with the device, not a compute cycle.

\begin{figure}[t]
    \scalebox{0.73}{\hspace{0cm}%%%%%%%%%%%%%
%  FIG 5
%%%%%%%%%%%%%

 \begin{minipage}[c]{1.0\linewidth}
  %\centering
  \[
   \begin{array}{cccccc}
      \begin{tikzpicture}[font=\large]
   
   \draw[style=dashed] (2,0) -- (4,0);

\node[draw] (p) at (3,-1) {$\p$};
\node[draw] (mp) at (3,2) {$m_\p$};
\draw[->,double] (mp) -- (p);
\draw (3.8,0.5) node {$\IRT$};

\node[draw] (ms) at (1,2.5) {$m_\mathbf{s}$};
 
\draw[->] (ms) to[bend left=30] (mp); 
\draw (2.25,2.95) node {$\Delta$}; 
   
   \end{tikzpicture}
     & \qquad \qquad     &
     
   \begin{tikzpicture}[font=\large]
   
   \draw[style=dashed] (2,0) -- (8,0);
   
   \node[draw] (p) at (3,-1) {$\p$};
\node[draw] (mp) at (3,2) {$01,10$};
\draw[->,double] (mp) -- (p);
\draw (3.8,0.5) node {$\IRT$};

\node[draw] (mprp) at (7,2) {$11$};

\draw[-open triangle 45] (mp) -- (mprp);
\draw (4.8,2.5) node {$C_\theory (01,10)$};
\draw (4.8,1.6) node {(gates)};

\node[draw] (ppr) at (7,-1) {$\p^\prime$};
\draw[-open triangle 45] (p) -- (ppr);
\draw (5.25,-.75) node {$\mathbf{H}(\p)$};
\draw (5,-1.5) node {(voltage changes)};

\draw[->,double] (ppr) -- (mprp);
\draw (7.75,0.5) node {$\RT$};

\end{tikzpicture}
     & \qquad \qquad    &
      \begin{tikzpicture}[font=\large]
   
   \draw[style=dashed] (2,0) -- (8,0);

\node[draw] (p) at (3,-1) {$\p$};
\node[draw] (mp) at (3,2) {$m_\p$};
\draw[->,double] (mp) -- (p);
\draw (3.8,0.5) node {$\IRT$};

\node[draw] (mprp) at (7,2) {$ m_{\p^\prime} \approx m^\prime_\p$};

\draw[-open triangle 45] (p) -- (ppr);

\node[draw] (ppr) at (7,-1) {$\p^\prime$};

\draw[->,double] (ppr) -- (mprp);
\draw (7.75,0.5) node {$\RT$};

\end{tikzpicture}

         \\{}\\
(a) & \qquad \qquad  &  (b)& \qquad \qquad  &  (c)
 \end{array}
\]
\end{minipage}}
\caption{Physical computing: (a) Embedding an abstract problem $M(S)$ into an abstract machine description $m_\p$ using embedding $\Delta$, then encoding into $\p$. (b) Full AR diagram for addition of two binary numbers using a computer (compare with figure \ref{com}). (c) The `compute cycle': using a reversed representation relation to encode data, physical evolution of the computer is used to predict abstract evolution.}\label{computecy}
\end{figure}
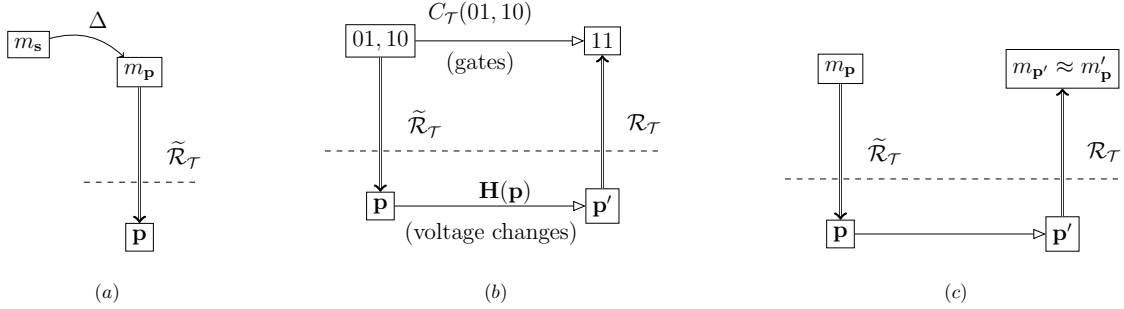

The compute cycle starts from a set of abstract objects -- the program and initial state that are to be computed. The existence of an abstract problem, embedded usually as an input and a program, is the reason why a physical computer is to be used, figure \ref{computecy}(a). Consider again the problem of figure \ref{com}, binary addition of two 2-bit numbers: $01 + 10 =11$. 

The abstract initial state, $m_\p = \{01,10\}$ is encoded, through $\IRT$, in the physical system $\p$, figure \ref{computecy}(a). The representation relation determines that detecting a high voltage corresponds to representing a `1', and low voltage is `0'. The initial physical setup therefore instantiates an initial abstract state. In this example, two parts of the hardware are designated by $\RT$ as `registers', and the voltages in the components of those areas correspond to the representation of the initial state as `01' and `10' (the two numbers we wish to add). 

At the abstract level, this initial state is now the input to a sequence of gate operations $C_\mathcal{T}$ that takes `01,10' and performs addition, figure \ref{computecy}(b) . At the physical level, a physical evolution $\mathbf{H}(\p)$ is applied to the state, producing the final physical state $\p^\prime$. Here, this will be the hardware manipulation of voltages. 

Finally, an application of $\RT$ takes the final physical state and represents it abstractly as some $m_{\p^\prime}$. After this decoding step, if the computer has the correct answer then $m^\prime_\p = (11)$. If we have confidence in the theory of the computer, then we are confident that $m_{\p^\prime} = m^\prime_\p$, and that this would be the outcome of the abstract evolution.

While a computer can be described in the above terms, as a parallel evolution of abstract and physical systems, the most important use of a computing system is when the abstract outcome $m'_\p$ is unknown: when computers are used to solve problems. In our example, if the outcome of $01+10$ were unknown, and the computing device being used to compute it, the final abstract state, $m^\prime_\p=(11)$, would not be evolved abstractly. Instead, confidence in the technological capabilities of the computer would enable the user to reach the final, abstract, output state $m_{\p^\prime} = m^\prime_\p$ using the physical evolution of the computing device alone. 

This use of a physical computer is the \textit{compute cycle}, figure \ref{computecy}(c): 
\[ (m_\p \xrightarrow{\IRT} \p \longrightarrow \p' \xrightarrow{\RT} [m_{\p'} = m'_{\p}])\]

\noindent This gives us the following definitions within AR theory:

\begin{definition} A \textbf{computer} is a (usually highly-engineered) device in the physical domain with a good and valid theory $\mathcal{T}$, together with representation and instantiation relations, $\RT,\IRT$, and the relevant commuting diagrams, which support compute cycles for its specific domain of inputs $\{ m_\p\}$ and computational operations $\{C_\mathcal{T}\}$.
\end{definition}

\begin{definition}
\textbf{Physical computing} is the use of a computer (a physical system) to predict the outcome of a computation (an abstract evolution) through a compute cycle. \label{defcomp}\end{definition}

%3 steps
%compute cycle p - mp - mpp
%prediction device2
%\footnote{I have been narrowly dissuaded from describing this conclusion ``no computation without representation".}

\section{Refinement and representation}

The definition of physical computing, Definition \ref{defcomp}, makes clear what was noted at the beginning of this paper: that AR theory is a \textit{supplement} to existing theories of computation, rather than a replacement of them. With the main machinery of AR theory in place, we can now see exactly where this framework fits in terms of the standard formal methods of computer science.

Perhaps the most immediate similarity to existing methods is with those of \textit{refinement theory} (see for example \cite{refine}). Refinement (alternatively, \textit{reification}), in its simplest form, takes an abstract algorithm and produces an equivalent concrete machine code that can be implemented on a computer. In general, it describes the relationship between different \textit{levels of abstraction} in computing code. For a refinement to be correct, the more concrete representation must faithfully implement the abstract specification. The concrete and abstract semantics of Abstract Interpretation, discussed briefly above, can be viewed as related by a refinement.

Refinement between levels of abstraction is governed by logico-mathematical refinement relations. If these relations are correct, then the levels of abstraction form commuting diagrams. At first sight, these refinement relations seem very similar to the representation and instantiation relations, and we will see that there are connections between the two. However, as with Abstract Interpretation, the operations and outcomes of refinement take place entirely within the abstract domain: even once a refinement is complete down to a machine code, this is still an \textit{abstract} specification of the higher-level algorithm.

\begin{figure}[t]
    \scalebox{0.65}{\hspace{0cm}%%%%%%%%%%%%%
%  FIG 6
%%%%%%%%%%%%%

 \begin{minipage}[c]{1.0\linewidth}
  %\centering
  \[
   \begin{array}{ccccc}
   \begin{tikzpicture}[font=\large]
   
   \draw[style=dashed] (2,0) -- (9,0);
   
   \node[draw] (p) at (3,-1) {$\p(asm)$};
\node[draw] (mp) at (3,2) {$01,10$};
\draw[->,double] (mp) -- (p);

\node[draw] (mprp) at (7,2) {$11$};

\draw[-open triangle 45] (mp) -- (mprp);

\node[draw] (ppr) at (7,-1) {$\p^\prime(asm)$};
\draw[-open triangle 45] (p) -- (ppr);
\draw (5,-.75) node {$\mathbf{H}(asm)$};

\draw[->,double] (ppr) -- (mprp);
\draw (8.2,0.5) node {$\mathcal{R}_{\theory(asm)}$};
\draw (4.1,0.5) node {$\tilde{\mathcal{R}}_{\theory(asm)}$};
\draw (5,2.4) node {$asm \ add$};

%%%layer 2 of refinement
\node[draw] (mp1) at (3,4) {$01,10$};
\node[draw] (mprp1) at (7,4) {$11$};

\draw[->] (mp1) -- (mprp1);
\draw[<-] (mp) -- (mp1);
\draw[->] (mprp) -- (mprp1);

\draw (7.5,3) node {$S_{BC}$};
\draw (3.5,3) node {$S_{BC}$};
\draw (5,4.4) node {$binary \ add$};

%%%layer 3 of refinement
\node[draw] (mp2) at (3,6) {$1,2$};
\node[draw] (mprp2) at (7,6) {$3$};

\draw[->] (mp2) -- (mprp2);
\draw[<-] (mp1) -- (mp2);
\draw[->] (mprp1) -- (mprp2);

\draw (7.5,5) node {$S_{AB}$};
\draw (3.5,5) node {$S_{AB}$};
\draw (5,6.4) node {$dec \ add$};

\end{tikzpicture}
%%%%%%%%
     &    &
%%%%%%%%
      \begin{tikzpicture}[font=\large]
   \draw[style=dashed] (2,0) -- (9,0);
   
   \node[draw] (p) at (3,-1) {$P(bin)$};

\draw[->,double] (mp1) -- (p);

\node[draw] (ppr) at (7,-1) {$\p^\prime(bin)$};
\draw[-open triangle 45] (p) -- (ppr);
\draw (5,-.75) node {$\mathbf{H}(bin)$};

\draw[->,double] (ppr) -- (mprp1);
\draw (8.2,0.5) node {$\mathcal{R}_{\theory(bin)}$};
\draw (4.1,0.5) node {$\tilde{\mathcal{R}}_{\theory(bin)}$};

%%%layer 2 of refinement
\node[draw] (mp1) at (3,4) {$01,10$};
\node[draw] (mprp1) at (7,4) {$11$};

\draw[->] (mp1) -- (mprp1);

\draw (5,4.4) node {$binary \ add$};

%%%layer 3 of refinement
\node[draw] (mp2) at (3,6) {$1,2$};
\node[draw] (mprp2) at (7,6) {$3$};

\draw[->] (mp2) -- (mprp2);
\draw[<-] (mp1) -- (mp2);
\draw[->] (mprp1) -- (mprp2);

\draw (7.5,5) node {$S_{AB}$};
\draw (3.5,5) node {$S_{AB}$};
\draw (5,6.4) node {$dec \ add$};
 
\end{tikzpicture}
%%%%%%%%
  &    &
%%%%%%%%
      \begin{tikzpicture}[font=\large]
   
      \draw[style=dashed] (2,0) -- (9,0);
   
   \node[draw] (p) at (3,-1) {$P(dec)$};

\draw[->,double] (mp2) -- (p);

\node[draw] (ppr) at (7,-1) {$\p^\prime(dec)$};
\draw[-open triangle 45] (p) -- (ppr);
\draw (5,-.75) node {$\mathbf{H}(dec)$};

\draw[->,double] (ppr) -- (mprp2);
\draw (8.2,0.5) node {$\mathcal{R}_{\theory(dec)}$};
\draw (4.1,0.5) node {$\tilde{\mathcal{R}}_{\theory(dec)}$};

%%%layer 3 of refinement
\node[draw] (mp2) at (3,6) {$1,2$};
\node[draw] (mprp2) at (7,6) {$3$};

\draw[->] (mp2) -- (mprp2);

\draw (5,6.4) node {$dec \ add$};
 
\end{tikzpicture}
         \\{}\\
(a) &     &  (b) &      &  (c)
 \end{array}
\]
\end{minipage}}
\caption{\label{refn} Physical computation, with layers of refinement $S$ on top for base ten (decimal) addition (``$dec \ add$"), binary addition (``$binary \ add$"), and assembly language addition (``$asm \ add$"). Note the physical device and representation differ in each case.}
\end{figure}
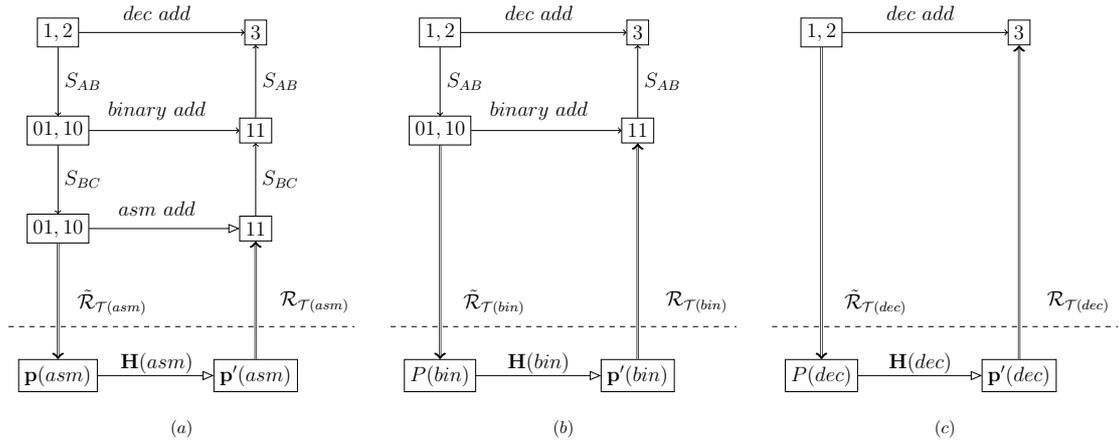

AR theory sits underneath this bottom level of a refinement stack, connecting the abstract system specification with the physical computing device. Figure \ref{refn} shows the connected stack reaching into the physical domain via the compute cycle. Different machine codes and different physical computers are given in each example. As a consequence, the representation and instantiation relations in each example are different.

The connection between representation and refinement now becomes clear: a representation can be viewed as similar to a refinement between physical and abstract levels. $\RT$ and $\IRT$ differ from $S_{AB,BC}$ in being non-mathematical relations, and in not being bi-directional; however, there are important structural similarities between AR theory and refinement theory that can now be brought out.

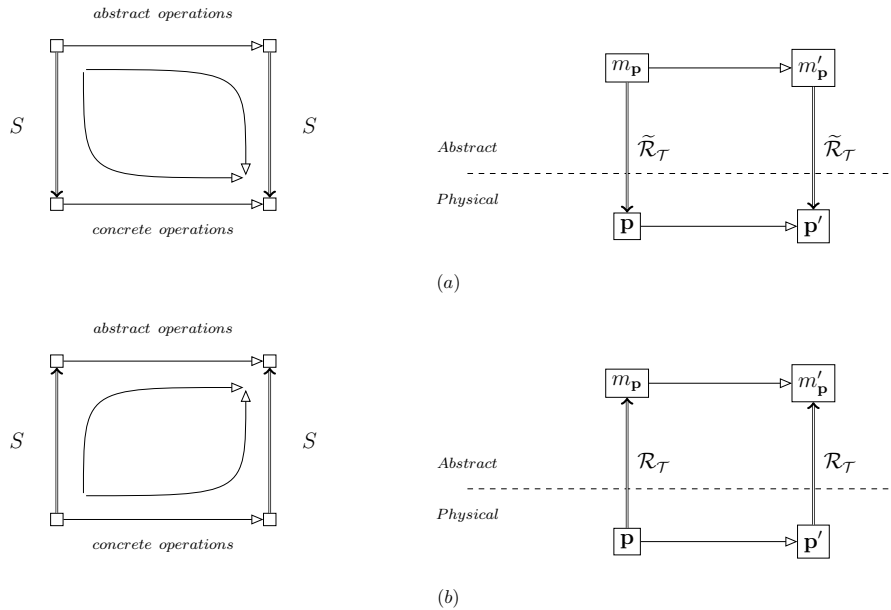
\begin{figure}[t]
    \scalebox{0.7}{\hspace{2cm}%%%%%%%%%%%%%
%  FIG 5
%%%%%%%%%%%%%

 \begin{minipage}[c]{1.0\linewidth}
  %\centering
  \[
   \begin{array}{cccccc}

   \begin{tikzpicture}[font=\large]
   
 %  \draw[style=dashed] (2,0) -- (8,0);
   
   \node[draw] (p) at (3,-1) {};
\node[draw] (mp) at (3,2) {};
\draw[->,double] (mp) -- (p);
\draw (2.25,0.5) node {$S$};

\node[draw] (mprp) at (7,2) {$$};

\draw[-open triangle 45] (mp) -- (mprp);
\draw (4.8,2.5) node {$$};
\draw (5,2.6) node {$\scriptstyle{abstract \ operations}$};

\node[draw] (ppr) at (7,-1) {$$};
\draw[-open triangle 45] (p) -- (ppr);
\draw (5.25,-.75) node {$$};
\draw (5,-1.5) node {$\scriptstyle{concrete \ operations}$};

\draw[->,double] (mprp) -- (ppr);
\draw (7.75,0.5) node {$S$};

%\draw[->,double] (3.5,1.5) to[out=270,in=180] (6,-0.5);
%\draw[->,double] (3.6,1.5) to[out=45,in=45] (6.1,-0.5);

\draw[-open triangle 45] (3.5,1.5) .. controls (3.5,-0.5) .. (6.5,-0.5);
\draw[-open triangle 45] (3.55,1.55) .. controls (6.55,1.55) .. (6.55,-0.45);

\end{tikzpicture}

\hspace{2cm}

   \begin{tikzpicture}[font=\large]
   
\draw[style=dashed] (0,0) -- (8,0);
\draw (0,0.5) node {$\scriptstyle{Abstract}$};
\draw (0,-0.5) node {$\scriptstyle{Physical}$};

%fig a
\node[draw] (p) at (3,-1) {$\mathbf{p}$};
\node[draw] (mp) at (3,2) {$m_{\mathbf{p}}$};
\draw[->,double] (mp) -- (p);
\draw (3.5,0.5) node {$\IRT$};

%fig b
\node[draw] (mprp) at (6.5,2) {$m^\prime_{\mathbf{p}}$};
\draw[-open triangle 45] (mp) -- (mprp);
%\draw (4.8,2.5) node {$C_\theory(m_\mathbf{p})$};

%fig c
\node[draw] (ppr) at (6.5,-1) {$\mathbf{p}^\prime$};
\draw[-open triangle 45] (p) -- (ppr);
%\draw (5.25,-.75) node {$\mathbf{H}(\mathbf{p})$};

%fig d
%\node[draw] (mppr) at (8,1) {$m_{\mathbf{p}^\prime}$};
\draw[->,double] (mprp) -- (ppr);
\draw (7,0.5) node {$\IRT$};

\end{tikzpicture}
\\{}\\
(a)
\\{}\\
        \begin{tikzpicture}[font=\large]
   \node[draw] (p) at (3,-1) {};
\node[draw] (mp) at (3,2) {};
\draw[->,double] (p) -- (mp);
\draw (2.25,0.5) node {$S$};

\node[draw] (mprp) at (7,2) {$$};

\draw[-open triangle 45] (mp) -- (mprp);
\draw (4.8,2.5) node {$$};
\draw (5,2.6) node {$\scriptstyle{abstract \ operations}$};

\node[draw] (ppr) at (7,-1) {$$};
\draw[-open triangle 45] (p) -- (ppr);
\draw (5.25,-.75) node {$$};
\draw (5,-1.5) node {$\scriptstyle{concrete \ operations}$};

\draw[->,double] (ppr) -- (mprp);
\draw (7.75,0.5) node {$S$};

\draw[-open triangle 45] (3.5,-0.5) .. controls (3.5,1.5) .. (6.5,1.5);
\draw[-open triangle 45] (3.55,-0.55) .. controls (6.55,-0.55) .. (6.55,1.45);

\end{tikzpicture}

\hspace{2cm}

   \begin{tikzpicture}[font=\large]
   
\draw[style=dashed] (0,0) -- (8,0);
\draw (0,0.5) node {$\scriptstyle{Abstract}$};
\draw (0,-0.5) node {$\scriptstyle{Physical}$};

%fig a
\node[draw] (p) at (3,-1) {$\mathbf{p}$};
\node[draw] (mp) at (3,2) {$m_{\mathbf{p}}$};
\draw[->,double] (p) -- (mp);
\draw (3.5,0.5) node {$\RT$};

%fig b
\node[draw] (mprp) at (6.5,2) {$m^\prime_{\mathbf{p}}$};
\draw[-open triangle 45] (mp) -- (mprp);
%\draw (4.8,2.5) node {$C_\theory(m_\mathbf{p})$};

%fig c
\node[draw] (ppr) at (6.5,-1) {$\mathbf{p}^\prime$};
\draw[-open triangle 45] (p) -- (ppr);
%\draw (5.25,-.75) node {$\mathbf{H}(\mathbf{p})$};

%fig d
%\node[draw] (mppr) at (8,1) {$m_{\mathbf{p}^\prime}$};
\draw[->,double] (ppr) -- (mprp);
\draw (7,0.5) node {$\RT$};

\end{tikzpicture}
\\{}\\
 (b)
 \end{array}
\]
\end{minipage}}
\caption{\label{simu} Analogues between cycles in AR and refinement theories: (a) Downwards/forwards simulation (also called history) and technology; (b) Upwards/backwards simulation (also called prophecy) and scientific prediction.  }
\end{figure}

The full use of a refinement for computation, shown in the upper levels of figure \ref{refn}, is verified through the correctness of one of two separate cycles: those of \textit{downwards/forwards} and \textit{upwards/backwards} simulation \cite{forwardbackwards} (for example). The use of forward vs backward refinement proofs is governed by the point in the computation that any indeterminism is resolved \cite{susanetal}. These cycle are correct if the corresponding diagrams commute. Replacing a downwards refinement arrow with $\RT$ and an upwards one with $\IRT$, we can give the AR diagrams spanning the abstract-physical divide which correspond to these two types of simulation, figure \ref{simu}.

We see immediately that \textit{downwards simulation is analogous to technology} (implementing an abstract specification into a physical device) and that \textit{upwards simulation is analogous to scientific prediction} (determining an abstract prediction for a physical event). The alternative terms for these simulations, `history' and `prophecy' respectively, import directly into the AR framework and become extremely descriptive \cite{historyprophecy}. It is a fascinating implication of AR theory that the relationship between abstract and physical objects follows the analogy with levels of abstraction in refinement theory even \textit{outside} the domain of computer science. One interesting point to note is that, for refinement, only one of upwards and downwards simulation is needed to prove the refinement. By contrast, AR theory requires both its analogous cycles for computing: physical computing is the combination of experimental device theories capable of prediction (equivalent to upwards simulation/prophecy), and of technological engineering of the device (equivalent to downwards simulation/history). In other words, a computer is a combination of scientific understanding and of technology use.

An interesting area for further investigation, seen also by the analogues with refinement, is the category-theory structure of the relationship between abstract and physical objects. There is an obvious categorical underpinning to AR theory in its commuting-diagrammatic form, and an interesting question arises as to whether representation, while itself not mathematical, can itself be represented in this way.

%this is all ref the Jonson Naumann Power paper.
One categorical account of data refinement was given by Hoare, representing refinement as a lax natural transformation between two models of the same category \cite{hoare}. If the computational system is embedded in a category $\mathcal{C}$, and $M,N$ are two models of this category, then if there is a lax natural transformation $\alpha: M \rightarrow N$ then $M$ is a `representation' of $N$ which is `more defined' (that is, the functors of $M$ are defined over a larger number of objects than the functors of $N$). Lax transformation is then the categorical representation of downwards/forwards simulation. More higher-level generalisations have also been studied, with the objective of finding a better categorical description of refinement. Such notions as lax \textit{logical} transformation \cite{laxlogical} point towards a way in which the similar concept of representation may be categorified, with a physical object and its own abstract representation given as generalised models of a single categorical structure.

%CATEGORICAL STUFF HERE Hoare, Power, natural/lax transformations etc

% with the machinery of ART, how do we add to the foundations of CS?

%AR theory supplements formal methods does not replace.

% abstraction - ie going from physical to abstract - as a form of upward simulation (in the extended refinement with AR structure underneath). Downward simulation = lax transformation. Note: no, abstraction is DOWNWARDS simulation: because the representation is of a smaller complexity than the entity itself. So the complexity of the abstraction meets at the lowest ABSTRACT level of the refinement-plus-ar stack.
% A REPRESENTATION IS A REFINEMENT

%Say: there is an obvious categorical underpinning to AR theory. Reference Hoare and Power. Come back to me in a few years' time.

% Refinement and categorical structure: overall system/language is embedded in category C. M and N are MODELS of this category which preserve the structure. M and N are related by (lax) natural transformations -> M is a refinement of N. M is "more defined" than N - its functors are defined over a larger number of objects than those of N.

% Matty's comment: can we say that a Turing Machine is a good model for some p? How can we say it? Is this formally undecidable?

\section{Meaning and ontology}
% models of formal systems vs representations of physical entities
% l-calc, Jamie's slides, distinguishing Lx -> x as a physical rep and as a formal system model
% analogy with the development of theoretical physics

A common issue in the foundations of computing concerns the meaning of objects within the formal systems of computation. By connecting these formal methods to the physical domain, AR theory helps refine questions of meaning and reference within computer science.

One of the most fundamental formal models of computation is that of the \textit{lambda calculus}. In itself, it is a formal system of symbol manipulation. It has interpretations and models as anything from the theory of functions, proofs, or programming languages, to definitions of effective computability, and the internal mechanics of cartesian closed categories. Depending on which model is being used, the semantic content to even the simple term $\lambda x . x$ varies enormously.

With an AR underpinning to the abstract domain, it is now possible to distinguish between a mathematical object $x$ being the representation of a physical system, and that same object belonging to a model of a $\lambda$-calculus. The set of abstract objects that are representations of physical objects, and the set of abstract objects representing models of formal systems, are not co-extensive. Some formal systems may have a concrete model but no physical referent; and some physical systems may have a mathematical abstraction that does not correspond to a model for any current formal system of computation.

Physical computing ocurrs in the special case where the abstract representation of the computing system, `bottom-up', is \textit{also} an object within a model of computation, `top-down'. In this case the semantic content to a computational assertion is two-fold, and the reference of the abstract object is similarly dual: the meaning of a term, say $\lambda yx . xy$ is both a computational assertion and a description of the physical capabilities of the physical computing device (here, that the state of two data-instantiating systems can be swapped). While formal computation theory extends beyond objects that are representations of physical devices, it is based around, and must be consistent with, the core of objects and relations that are. 

This dual ontology of mathematical objects, both physical and abstract, while unusual in the context of computer science, is the standard within the natural sciences. A comparison with the situation in theoretical physics is highly instructive here. Physics started bottom-up, with many physical systems which were then represented abstractly, and the formal mathematics of their representations developed. Sometimes it was noted that the mathematical representations of these physical systems also corresponded to elements of previously known mathematics. 

By contrast, computer science has historically concentrated on formal systems, top-down, with only a very narrow range of physical systems corresponding to the abstract specifications. The challenge for non-standard and extended types of physical computing is to import some of the methodology of natural science into computer science. Just as theoretical physics is now starting to consider formal systems that may not correspond to physical universes (for example generalised probabilistic theories based on quantum mechanics \cite{jongpt}), so computer science needs to extend to deal with physical systems that do not immediately correspond with standard formal methods. Abstract representations of non-standard computing systems may be found to correspond through some embedding or refinement or change in representation with the abstract representation of standard computers. In some cases they may, however, require an abstract theory of computation that is different; or another instantiation of a more general theoretical framework. The framework of AR theory can help allow us to make that determination for novel computing devices.

This leads to what is not a definition, but a description of the new addition AR theory gives computer science, in the form of physical computing:
\begin{quote} \textbf{Computing} is the natural science of the computational abilities of physical systems. \end{quote}

\section{Heterotic and hybrid computing}

We now apply AR theory and our understandings of refinement and ontologies to the central issue of this paper: the representation of heterotic and hybrid computing within AR theory.

In order to highlight the central feature of heterotic computing, we first make a distinction with non-heterotic hybrid computing. As noted in the introduction, heterotic hybrid computing allows computations to be performed that are, in some sense to be determined, more powerful than, or otherwise significantly different from, the computations that can be run on the individual systems. By contrast, we will use the term `hybrid', without modifier, where the computations on the joint system are a simple composition of the computations on the individual ones. 

A simple example of a hybrid system would be the human-computer interaction when, in order to perform the calculation $(10+10)/7$ the human user mentally performs $(10+10)$ and then uses a computer to calculate $20/7$. The joint computation is hybrid, but is not outside the computing power of either individually. Contrast this with another human-computer interaction, a social machine: for a machine such as Galaxy Zoo \cite{galaxy, galaxypaper}, the computation performed is outside the abilities of either the humans or the computer: the machine is heterotic.

We will postulate here a type of hybrid-heterotic distinction that is native to AR theory. It may be \textit{the} distinction, or it may be one type of a class of distinctions given by different types of composition; this, and the relationship to heterotic systems as defined by the original authors \cite{SS-UC11,susanvivspecis}, are areas for further investigation. The following can all be thought of as a consideration of \textit{representational} hybrid vs. heterotic systems.

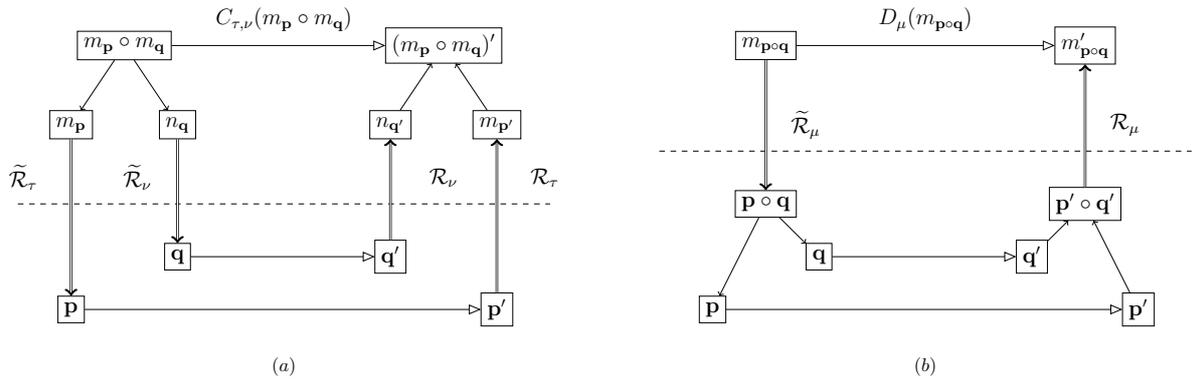
\begin{figure}[t]
    \scalebox{0.7}{\hspace{-1cm}%%%%%%%%%%%%%
%  FIG 2
%%%%%%%%%%%%%

 \begin{minipage}[c]{1.0\linewidth}
  %\centering
  \[
   \begin{array}{ccc}
   \begin{tikzpicture}[font=\large]
   
   \draw[style=dashed] (0,0) -- (10,0);
   
   \node[draw] (s) at (3,-1) {$\mathbf{q}$};
   \node[draw] (p) at (1,-2) {$\p$};
   \node[draw] (spr) at (7,-1) {$\mathbf{q}^\prime$};
   \node[draw] (ppr) at (9,-2) {$\p^\prime$};
   
   \node[draw] (ns) at (3,1.5) {$n_\mathbf{q}$};
   \node[draw] (mp) at (1,1.5) {$m_\p$};
   \node[draw] (nspr) at (7,1.5) {$n_{\mathbf{q}^\prime}$};
 %  \node[draw] (mppr) at (6.3,2.5) {$m^\prime_\p$};   
   \node[draw] (mprp) at (9,1.5) {$m_{\p^\prime}$};
   \node[draw] (mon) at (2,3) {$m_\p \circ m_\mathbf{q}$};
   \node[draw] (monpr) at (8,3) {$(m_\p \circ m_\mathbf{q})'$};
   
   \draw[-open triangle 45] (s) -- (spr);
   \draw[-open triangle 45] (p) -- (ppr);
   \draw[->,double] (ns) -- (s);
   \draw[->,double] (mp) -- (p);
   \draw[->,double] (spr) -- (nspr);
   \draw[->,double] (ppr) -- (mprp);
   
   \draw[-open triangle 45] (mon) -- (monpr);
   
   \draw[->] (mon) -- (mp);
  \draw[->] (mon) -- (ns);
  \draw[->] (nspr) -- (monpr);
  \draw[->] (mprp) -- (monpr);
   
   \draw (0.1,0.5) node {$\mathcal{\widetilde{R}_{\tau}}$};
   \draw (9.9,0.5) node {$\mathcal{R}_{\tau}$};
   \draw (2.25,0.5) node {$\mathcal{\widetilde{R}_{\nu}}$};
   \draw (8,0.5) node {$\mathcal{R}_{\nu}$};
   
   \draw (5,3.5) node {$C_{\tau ,\nu}(m_\p \circ m_\mathbf{q})$};
  
   \end{tikzpicture}
& \qquad  
\quad &   

    \begin{tikzpicture}[font=\large]
   
   \draw[style=dashed] (0,1) -- (10,1);
   
   \node[draw] (s) at (3,-1) {$\mathbf{q}$};
   \node[draw] (p) at (1,-2) {$\p$};
   \node[draw] (spr) at (7,-1) {$\mathbf{q}^\prime$};
   \node[draw] (ppr) at (9,-2) {$\p^\prime$};
   
 %  \node[draw] (ns) at (3,1.5) {$n_\mathbf{q}$};
  % \node[draw] (mp) at (1,1.5) {$m_\p$};
  % \node[draw] (nspr) at (7,1.5) {$n_{\mathbf{q}^\prime}$};
 %  \node[draw] (mppr) at (6.3,2.5) {$m^\prime_\p$};   
  % \node[draw] (mprp) at (9,1.5) {$m_{\p^\prime}$};
  
  \node[draw] (poq) at (2,0) {$\p \circ \mathbf{q} $};
    \node[draw] (poqpr) at (8,0) {$\p' \circ \mathbf{q'} $};
   \node[draw] (mon) at (2,3) {$m_{\p \circ \mathbf{q}}$};
   \node[draw] (monpr) at (8,3) {$m'_{\p \circ \mathbf{q}}$};
   
   \draw[-open triangle 45] (s) -- (spr);
   \draw[-open triangle 45] (p) -- (ppr);
   \draw[->] (poq) -- (p);
   \draw[->] (poq) -- (s);
   \draw[->] (ppr) -- (poqpr);
   \draw[->] (spr) -- (poqpr);
   \draw[->,double] (mon) -- (poq);
   \draw[->,double] (poqpr) -- (monpr);
  
%   \draw[->,double] (mon) .. controls (2.25,0) .. (p);
 %    \draw[->,double] (mon) .. controls (2.25,0) .. (s);
     
 %    \draw[->,double] (spr) .. controls (8.25,0) .. (monpr);
  %   \draw[->,double] (ppr) .. controls (8.25,0) .. (monpr);

   \draw[-open triangle 45] (mon) -- (monpr);
   
%   \draw[->] (mp) -- (mon);
 % \draw[->] (ns) -- (mon);
  %\draw[->] (nspr) -- (monpr);
 % \draw[->] (mprp) -- (monpr);
   
   \draw (8.75,1.5) node {$\mathcal{R}_{\mu}$};
   \draw (2.75,1.5) node {$\mathcal{\widetilde{R}_{\mu}}$};
   
   \draw (5,3.5) node {$D_{\mu}(m_{\p \circ \mathbf{q}} ) $};
  
   \end{tikzpicture}   \\{}\\
(a) & \qquad \quad  &  (b)
 \end{array}
\]
\end{minipage}}
\caption{Commuting physical and abstract diagrams for (a) hybrid vs (b) heterotic computing.}\label{hetero}
\end{figure}

A hybrid (unmodified) system in AR theory can be described using composition of computation in the abstract domain. Figure \ref{hetero}(a) shows this for two physical computing devices, $\mathbf{p}$ and $\mathbf{q}$ (these could be a human and a computer respectively). They each encode and process information according to their own representation relations and theory. That is, the computational system forms two separate representational triples, $\langle \p , \mathcal{R_{\tau}} , m_\p \rangle$ and $\langle \mathbf{q} , \mathcal{R_{ \nu}} , m_\mathbf{q} \rangle$, where $\tau,\nu$ are the theories of the two computing systems.

There are now two possibilities for hybrid data processing. The first, as shown in figure \ref{hetero}(a), is parallel composition: the computation performed, $C$, itself a composition of $C_\tau, C_\nu$, is performed on the composed input:
\[ C_{\tau ,\nu}(m_\p \circ m_\mathbf{q}): m_\p \circ m_\mathbf{q} \longrightarrow (m_\p \circ m_\mathbf{q})' \]

\noindent The second option is sequential composition. This is where the computations $C_\tau, C_\nu$ are performed on separate machines and then composed:
\[ C_\tau(m_\p) \circ C_\nu(m_\mathbf{q}): m_\p \circ m_\mathbf{q} \longrightarrow m'_\p \circ m'_\mathbf{q} \]

The case of heterotic computing differs in significant respects from straightforward hybrid computing. In the case of heterotic systems, the composition happens in the representation itself, before the abstract computation is reached. Figure \ref{hetero}(b) shows the production of a joint representation for the heterotic system composed of $\p$ and $\mathbf{q}$. Importantly, there is only one representational triple formed for the joint system: 
\[ \langle \p \circ \mathbf{q} , \mathcal{R_{\mu }} ,  m_{\p \circ \mathbf{q}} \rangle \] 

\noindent The single representational triple then means that there is a single set of operations allowed on the abstract input, given by the composed device theory $\mu$. We will denote these operations $D_{\mu}$. The system then proceeds as
\[ D_{\mu}(m_{\p \circ \mathbf{q}} ) : m_{\p \circ \mathbf{q}}  \longrightarrow m'_{\p \circ \mathbf{q}} \]

The crucial difference in this heterotic case is that $D_{\mu}$ need not be a composition of $C_\tau , C_\nu$. Because the representation is combined, the joint system can have access to computational operations not available to individual ones. This is the signature of heterotic computing: that composition ocurrs in the representation, rather than within the abstract domain.

It will, in general, be delicate work to pick out exactly where the composition is occurring, and so to argue whether a system is `truly' heterotic or not. If the systems $\p$ and $\mathbf{q}$ are individually capable of computing, then they will participate in representational triples $\langle \p , \mathcal{R_{ \tau}} , m_\p \rangle$ etc individually, as well as jointly for the heterotic computation. Differentiating these two levels of representation will be important.

As an example, we will consider the Galaxy Zoo social machine. A human user interacts by classifying objects presented to them into one of several categories. The computer processes these inputs according to its programming and produces an output. The user can be described using $\langle \p , \mathcal{R_{ \tau}} , m_\p \rangle$, where $\tau$ is the computational theory of how a human computing device classifies visual objects as `round', `square', etc. The computer satisfies the triple $\langle \mathbf{q} , \mathcal{R_{ \nu}} , m_\mathbf{q} \rangle$, where $\nu$ is the theory of computers running Javascript. Individually, these computations make sense, and satisfy all the relevant commuting diagrams. 

However, a full description of the system is not reducible to a description in terms of these two elements. The social machine itself is described by the triple $ \langle \p \circ \mathbf{q} , \mathcal{R_{\mu }} ,  m_{\p \circ \mathbf{q}}\rangle$. The theory of this machine, $\mu$, is the theory of a computing device that takes as input astronomical pictures and outputs classifications of galaxies. Only in the context of theory $\mu$ are the individual actions of the human and the computer describable as part of this computation. Furthermore, this computation is not available to either individually (given our current understanding of the abilities of human beings and Javascript programs).

It is key to heterotic computing that the heterotic triple not be decomposable. For some composition of representational triples, then, we can give a the following definition:
\begin{definition} If two physical systems $\p$ and $\mathbf{q}$ are elements of the representational triples $\langle \p , \mathcal{R_{ \tau}} , m_\p \rangle$ and $\langle \mathbf{q} , \mathcal{R_{ \nu}} , m_\mathbf{q} \rangle$, and individually satisfy all the requirements for being a computer, and if the joint physical system $\p \circ \mathbf{q}$ also participates in the representational triple $ \langle \p \circ \mathbf{q} , \mathcal{R_{\mu }} ,  m_{\p \circ \mathbf{q}}\rangle$ and satisfies the requirements for being a computer, if
\[  \langle \p \circ \mathbf{q} , \mathcal{R_{\mu }} ,  m_{\p \circ \mathbf{q}}\rangle \neq  \langle \p , \mathcal{R_{ \tau}} , m_\p \rangle \circ \langle \mathbf{q} , \mathcal{R_{ \nu}} , m_\mathbf{q} \rangle \]

\noindent then the computing system is \textbf{heterotic}. If the equality holds, then the computing system is \textbf{hybrid}. \end{definition}

\noindent Defining composition between representational triples is an important area of further work.

This consideration of hybrid and heterotic brings us back to the centrality of representation in computing: the different abilities of heterotic systems over their hybrid cousins lie in their abilities to use composed representations of joint systems. We can also see another aspect of AR theory: that more than one representation can be given for the same physical system -- even more than one \textit{computational} representation, where the same physical system is taking part in more than one computation under different representational theories. By foregrounding representation, AR theory has given us the ability to distinguish these different computational scenarios; and so to reason rigorously about these novel computing devices and processes.

%THERE'S GOING TO BE A LOT MORE HERE ABOUT COMPOSITION OF THE PHYSICAL SYSTEM AND HOW THAT RELATES TO THIS ABSTRACT COMPOSITION VIA PREVIOUSLY DISCOVERED SETS OF COMMUTING DIAGRAMS 

%need a COMBINED theory: we can see that form social machines, that it's the theory that when humans get joined together that they can do these extra computations
% one signature of a heterotic system is that the computations performed in isolation don't make any sense: the data representation is spread across the whole system, and cannot be access locally.

%it can be delicate work to pick out exactly where the composition is ocurring, and so to argue whether a system is 'truly' heterotic or not.

% We now distinguish between the two. Describe.

%the difference is composition in the representation vs composition in the abstract domain

% AR theory can distinguish the situations here through composition

% what you get out of hybrid is a composition within the abstract computing class. What you get out of hybrid is a composition that is then represented: the class of computations {mm,C(mom)} can be far larger than {mm,Cm o Cm}. THIS IS HETEROTIC. It depends crucially on representation

% hybrid vs heterotic for human-computer interaction. Shows that it depends on the rep not the computation - can have humans doing the computing separately, and computers, but it's the representation of their joint computation that gives the power of the social machine.

\section{Conclusion}

%In this paper we have given the theoretical underpinnings of the natural science of physical computing through the use of Abstraction/Representation theory. 
%Focussed in particular on hybrid and heterotic computational devices and systems.

In this paper we have seen how Abstraction/Representation theory enables us to add physical systems, both standard, unconventional, and hybrid/heterotic, into the foundational language of computer science. By making rigorous the \textit{representation relation} between physical and abstract objects, we have been able to distinguish between computing in the physical domain and computation in the abstract -- and to give set of relations between the two. The algebraic and commuting-diagram forms of AR theory can be used to show the similarities and differences between different stages of computer design and use -- from the initial experiments on a substrate, to its engineering, and finally its use as a computational device. This has led us to the AR definition of physical computing, whereby a computer is shown to be a physical device for predicting the outcome of an abstract evolution (a computation). 

Representation is key to computing: a computer is a machine that must use the representation relation as part of its functioning, as unless it is present within a compute cycle then no computing can be said to be ocurring in a physical device.\footnote{One may be tempted to render this conclusion as ``no computation without representation''.} We have seen how the representation relation extends the refinement and abstraction relations of Abstract Interpretation and refinement theory. With the addition of AR theory, refinement diagrams can be given that commute `all the way down' to the physical device (not simply the still-abstract machine code), and the connections between the physics of the substrate and the levels of abstraction they can support becomes clearer. We have seen furthermore the beginnings of a fascinating relationship between types of refinement simulation, and different cycles in AR theory corresponding to science and the use of technology. 

% we have seen how AR theory extends refinement, and how it relates to lambda calculus and gives a physical ontology for computation theory. 

We have defined physical computing and described the development of the natural science of computing. We have seen how AR theory fits in with formal abstract systems of computation such as the lambda calculus and gives a physical ontology for computation theory. This now opens up the way to using AR theory in the formalisation of computing using novel physical system, such as quantum devices and the newly-described set of internet-enabled social machines. We have seen how these hybrid and heterotic systems can be defined and distinguished using AR theory. The power inherent in the AR separation of abstract and physical domains, and representation, comes out strongly as we are now able to distinguish between composition of representations (hybrid) and composition within representation (heterotic). By no longer conflating the representation of a device with the device itself, we are able to determine when a single device supports multiple computational representations, and the relationships between them.

There are many, many open questions and problems here: the development of AR theory and its application to computing is still in its first stages. It has already shown itself to be a powerful tool for reasoning about computational systems and processes, and further work will only add to its range of application. One immediate avenue is the extension of the analysis given above for social machines. A full construction for a specific machine, giving the precise connection between the different levels of representation present, and the formal definition of composition between representational triples, is outstanding work.

Another open, and very interesting, question is how AR theory in general relates in detail to the categorical substructures of theoretical computer science. In particular, the connection between refinement types and the AR commutation cycles for science and technology lead to the possibility that a categorification of AR theory would give an algebra for experimental theory confirmation in science -- and, by the Curry-Howard correspondence, an associated logic. This would be a novel logical structure for theory choice -- opening up in turn the possibility of importing it back into computer science as an addition to theory discovery and decision algorithms in artificial intelligence.

% Wherever this goes in the future, the gap between theory and device that has been historically present in CS has now been filled. The final piece of the puzzle of computer science is now in place.
AR theory does not just add new rigorous tools for computing. The foundational gap between computational theory and physical device that has historically been present has now been bridged. Computers show themselves as machines that manipulate both physics and representation, and computing as a process that necessarily crosses the divide between physical and abstract. A missing piece of the puzzle in the foundations of computer science is now in place.

\section{Acknowledgements}

I would like to thank Viv Kendon and Susan Stepney for many and continuing discussions on AR theory. Thanks are also owing to two anonymous referees for very useful comments and suggestions.

\bibliographystyle{unsrt}
\bibliography{speciss}

\begin{thebibliography}{10}

\bibitem{SS-UC11}
Viv Kendon, Angelika Sebald, Susan Stepney, Matthias Bechmann, Peter Hines, and
  Robert~C. Wagner.
\newblock Heterotic computing.
\newblock In {\em Unconventional Computation 2011, Turku, Finland, June 2011},
  volume 6714 of {\em LNCS}, pages 113--124. Springer, 2011.

\bibitem{susanvivspecis}
Viv Kendon, Angelika Sebald, and Susan Stepney.
\newblock Heterotic computing: past, present, and future.
\newblock {\em Philosophical Transactions of the Royal Society}, SPECIAL ISSUE
  VOLUME(NUMBER), 2015.

\bibitem{Ladd2010}
T~D Ladd, F~Jelezko, R~Laflamme, Y~Nakamura, C~Monroe, and J~L O'Brien.
\newblock {Quantum computers}.
\newblock {\em Nature}, 464(7285):45--53, 2010.

\bibitem{blueprint}
Rodney Van~Meter and Clare Horsman.
\newblock A blueprint for building a quantum computer.
\newblock {\em Commun. ACM}, 56(10):84--93, 2013.

\bibitem{chem1}
L~Kuhnert, K~I Agladze, and V~I Krinsky.
\newblock Image processing using light-sensitive chemical waves.
\newblock {\em Nature}, 337(6204):244--247, 1989.

\bibitem{chem2}
A.~Adamatzky, B.~De~Lacy~Costello, and T.~Asai.
\newblock {\em Reaction-Diffusion Computers}.
\newblock Elsevier, 2005.

\bibitem{dna1}
L~M Adleman.
\newblock Molecular computation of solutions to combinatorial problems.
\newblock {\em Science}, 266(5187):1021--1024, 1994.

\bibitem{dna2}
Martyn Amos.
\newblock {\em Theoretical and Experimental DNA Computation}.
\newblock Springer, 2005.

\bibitem{ratonaplane}
Thomas~B DeMarse and Karl~P Dockendorf.
\newblock Adaptive flight control with living neuronal networks on
  microelectrode arrays.
\newblock In {\em Neural Networks, 2005. IJCNN'05. Proceedings. 2005 IEEE
  International Joint Conference on}, volume~3, pages 1548--1551. IEEE, 2005.

\bibitem{sociam1}
Paul~R Smart and Nigel~R Shadbolt.
\newblock Social machines.
\newblock In Mehdi Khosrow-Pour, editor, {\em Encyclopedia of Information
  Science and Technology}, pages 6855--6862. 2014.

\bibitem{sociam2}
Paul~R Smart, Elena Simperl, and Nigel Shadbolt.
\newblock A taxonomic framework for social machines.
\newblock In Daniele~Miorandi \emph{et al}, editor, {\em Social Collective
  Intelligence: Combining the Powers of Humans and Machines to Build a Smarter
  Society}, pages 51--85. 2014.

\bibitem{wiki}
Wikipedia.
\newblock \url{http://en.wikipedia.org}, accessed 03/2015.

\bibitem{galaxy}
Christopher~Lintott \emph{et al}.
\newblock Galaxy {Z}oo.
\newblock \url{http://www.galaxyzoo.org}, accessed 03/2015.

\bibitem{compute}
Clare Horsman, Susan Stepney, Rob Wagner, and Viv Kendon.
\newblock When does a physical system compute?
\newblock {\em Proceedings of the Royal Society of London A}, 470(20140182),
  2014.

\bibitem{absint}
Patrick Cousot and Radhia Cousot.
\newblock Abstract interpretation: a unified lattice model for static analysis
  of programs by construction or approximation of fixpoints.
\newblock In {\em Conference Record of the Fourth Annual ACM SIGPLAN-SIGACT
  Symposium on Principles of Programming Languages}, pages 238--252. ACM Press,
  New York, 1977.

\bibitem{refine}
Jifeng He, C.~A.~R Hoare, and J~W Sanders.
\newblock Data refinement refined (resume).
\newblock In Bernard Robinet and Reinhard Wilhelm, editors, {\em ESOP 86},
  volume 213 of {\em LNCS}, pages 187--196. Springer, 1986.

\bibitem{forwardbackwards}
Moshe Deutsch and Martin~C Henson.
\newblock An analysis of forward simulation data refinement.
\newblock In {\em ZB 2003: Formal Specification and Development in Z and B},
  pages 148--167. Springer, 2003.

\bibitem{susanetal}
John~A. Clark, Susan Stepney, and Howard Chivers.
\newblock Breaking the model: finalisation and a taxonomy of security attacks.
\newblock Technical Report YCS-2004-371, Department of Computer Science,
  University of York, 2004.

\bibitem{historyprophecy}
Nancy Lynch and Fritz Vaandrager.
\newblock Forward and backward simulations 1: untimed systems.
\newblock {\em Information and Computation}, 121:214--233, 1995.

\bibitem{hoare}
C.~T. Hoare.
\newblock Data refinement in a categorical setting.
\newblock \url{http://www.cs.ox.ac.uk/files/6099/H87a\%20-\%20Data.pdf},
  accessed 03/2015.

\bibitem{laxlogical}
Michael Johnson, David Naumann, and John Power.
\newblock Category theoretic models of data refinement.
\newblock {\em Electronic Notes in Computer Science}, 255:21--38, 2009.

\bibitem{jongpt}
Jonathan Barrett.
\newblock Information processing in generalized probabilistic theories.
\newblock {\em Physical Review A}, 75(3):032304, 2007.

\bibitem{galaxypaper}
Kate~Land \emph{et al}.
\newblock Galaxy {Z}oo: the large-scale spin statistics of spiral galaxies in
  the sloan digital sky survey.
\newblock {\em MNRAS}, 388(4):1686--1692, 2008.

\end{thebibliography}

\end{document}